\documentstyle{mn}
\input psfig.tex
\def\lsim{\lower.5ex\hbox{$\; \buildrel < \over \sim \;$}}
\def\gsim{\lower.5ex\hbox{$\; \buildrel > \over \sim \;$}}
\def \simeq{\lower.3ex\hbox{$\; \buildrel \sim \over - \;$}}
\def\ch{\lower-0.55ex\hbox{--}\kern-0.55em{\lower0.15ex\hbox{$h$}}}
\def\lh{\lower-0.55ex\hbox{--}\kern-0.55em{\lower0.15ex\hbox{$\lambda$}}}
\newif\ifAMStwofonts
\ifoldfss
  \ifCUPmtlplainloaded \else
    \NewTextAlphabet{textbfit} {cmbxti10} {}
    \NewTextAlphabet{textbfss} {cmssbx10} {}
    \NewMathAlphabet{mathbfit} {cmbxti10} {} % for math mode
    \NewMathAlphabet{mathbfss} {cmssbx10} {} %  "   "    "
  \fi
  \ifAMStwofonts
    \ifCUPmtlplainloaded \else
      \NewSymbolFont{upmath} {eurm10}
      \NewSymbolFont{AMSa} {msam10}
      \NewMathSymbol{\upi}     {0}{upmath}{19}
      \NewMathSymbol{\umu}     {0}{upmath}{16}
      \NewMathSymbol{\upartial}{0}{upmath}{40}
      \NewMathSymbol{\leqslant}{3}{AMSa}{36}
      \NewMathSymbol{\geqslant}{3}{AMSa}{3E}

       \let\le=\leqslant
       
    \fi
  \fi
\fi % End of OFSS
\ifnfssone
  \newmathalphabet{\mathit}
  \addtoversion{normal}{\mathit}{cmr}{m}{it}
  \addtoversion{bold}{\mathit}{cmr}{bx}{it}
  \newmathalphabet{\mathbfit} % math mode version of \textbfit{..}
  \addtoversion{normal}{\mathbfit}{cmr}{bx}{it}
  \addtoversion{bold}{\mathbfit}{cmr}{bx}{it}
  \newmathalphabet{\mathbfss} % math mode version of \textbfss{..}
  \addtoversion{normal}{\mathbfss}{cmss}{bx}{n}
  \addtoversion{bold}{\mathbfss}{cmss}{bx}{n}
  \ifAMStwofonts
    \ifCUPmtlplainloaded \else
      \UseAMStwoboldmath
      \makeatletter
      \new@mathgroup\upmath@group
      \define@mathgroup\mv@normal\upmath@group{eur}{m}{n}
      \define@mathgroup\mv@bold\upmath@group{eur}{b}{n}
      \edef\UPM{\hexnumber\upmath@group}
      \new@mathgroup\amsa@group
      \define@mathgroup\mv@normal\amsa@group{msa}{m}{n}
      \define@mathgroup\mv@bold\amsa@group{msa}{m}{n}
      \edef\AMSa{\hexnumber\amsa@group}
      \makeatother
      \mathchardef\upi="0\UPM19
      \mathchardef\umu="0\UPM16
      \mathchardef\upartial="0\UPM40
      \mathchardef\leqslant="3\AMSa36
      \mathchardef\geqslant="3\AMSa3E

       \let\le=\leqslant

    \fi
  \fi
\fi % End of NFSS release 1

\ifnfsstwo
  \DeclareMathAlphabet{\mathbfit}{OT1}{cmr}{bx}{it}
  \SetMathAlphabet\mathbfit{bold}{OT1}{cmr}{bx}{it}
  \DeclareMathAlphabet{\mathbfss}{OT1}{cmss}{bx}{n}
  \SetMathAlphabet\mathbfss{bold}{OT1}{cmss}{bx}{n}
  \ifAMStwofonts
    \ifCUPmtlplainloaded \else
      \DeclareSymbolFont{UPM}{U}{eur}{m}{n}
      \SetSymbolFont{UPM}{bold}{U}{eur}{b}{n}
      \DeclareSymbolFont{AMSa}{U}{msa}{m}{n}
      \DeclareMathSymbol{\upi}{0}{UPM}{"19}
      \DeclareMathSymbol{\umu}{0}{UPM}{"16}
      \DeclareMathSymbol{\upartial}{0}{UPM}{"40}
      \DeclareMathSymbol{\leqslant}{3}{AMSa}{"36}
      \DeclareMathSymbol{\geqslant}{3}{AMSa}{"3E}

       \let\le=\leqslant

    \fi
  \fi
\fi % End of NFSS release 2

\ifCUPmtlplainloaded \else
  \ifAMStwofonts \else % If no AMS fonts
    \def\upi{\pi}
    \def\umu{\mu}
    \def\upartial{\partial}
  \fi
\fi

\title{Behaviour of matter close to the event horizon}
\author[Tapas K. Das]
       {Tapas K. Das $^{1,2}$\\
$^1$ Division of Astronomy, Department of Physics and Astronomy,
University of California at Los
Angeles, Box 951562, Los Angeles, \\
CA 90095-1562, USA\\
$^2$ Institute of Geophysics and Planetary Physics, University of California at Los
Angeles, Box 951567, Los Angeles, CA 90095, USA\\
tapas@astro.ucla.edu}

\date{Accepted .
      Received ;
      in original form }
\pubyear{}
\begin{document}
\onecolumn
%\twocolumn

\maketitle

\begin{abstract}
\noindent
{Investigation of the behaviour of accreting matter close to the black hole
event
horizon is of fundamental importance in relativistic and high energy
astrophysics because they provide the key features of the diagnostic
spectra of the stellar mass and super-massive black holes.
In this paper, we examine the terminal behaviour of general relativistic
matter in multi-transonic,
advective black hole accretion discs. We compute, for the first time we believe, the values
of various dynamical and thermodynamic multi-transonic
flow variables {\it extremely close} ($\le$ 0.01 $r_g$) to
the event horizon and study the dependence of these variables on
fundamental accretion parameters. Our calculation is
useful for a better
understanding of Hawking radiation from acoustic black holes.}
\end{abstract}

\begin{keywords} 
Accretion, accretion discs --- black hole physics --- general relativity --- hydrodynamics
--- Hawking Radiation\\[0.25cm]
\noindent
%Astrophysical black holes, accretion, outflows, radiation drag
\end{keywords} 
\section{Introduction}
\noindent
Gravitational capture of surrounding fluid by massive
astrophysical objects is known as accretion. There remains a major difference between black hole
(BH) accretion and accretion onto other cosmic objects including neutron stars and
white dwarfs. For celestial bodies other than black holes, infall of matter terminates
either by a direct collision with the hard surface of the accretor or with the outer boundary of the 
magneto-sphere, resulting the luminosity through energy release 
from the surface. Whereas for black hole accretion, matter ultimately 
dives through the event horizon from where radiation is prohibited to escape according 
to the rule of classical general relativity (GR) and the  emergence of luminosity occurs
on the way towards the black hole event horizon. The efficiency
of accretion process may be thought as a measure of the
fractional conversion of gravitational binding energy of matter
to the emergent radiation and is considerably high for black
hole accretion  compared to accretion onto any other
astrophysical objects. Hence accretion onto classical astrophysical black holes
has
been recognized as a fundamental phenomena of increasing
importance in relativistic and high energy astrophysics
(Frank, King \& Raine 1992, hereafter FKR, Shapiro \& Teukolsky 1983). The
extraction of gravitational energy from the black hole accretion is believed to
power the energy generation mechanism of
X-ray  binaries and of the most luminous objects of the
Universe, the Quasars and active galactic nuclei (AGN). The
BH accretion is, thus, the most appealing way through which the
all pervading power of gravity is explicitly manifested.
If the infalling matter does not possess intrinsic angular momentum,
accretion flow remains spherically symmetric. However,
Inter stellar / intergalactic fluid is always likely to posses
non-vanishing rotational energy, sufficient to dynamically
break the spherical symmetry, and in almost all real physical
situations, accreting matter is thrown into circular orbits
around the central accretor, leading to the formation of the
{\it accretion disc} around the galactic and extra-galactic
black holes. \\
\noindent
If the instantaneous dynamical velocity and local acoustic velocity
 of the accreting fluid, moving along a space curve parameterized by $r$, are
$u(r)$ and $a(r)$ respectively, then the local Mach number $M(r)$ of the
 fluid can be defined as $M(r)=\frac{u(r)}{a(r)}$.
The flow will be locally
 subsonic or supersonic according to $M(r) < 1$ or $ >1$, i.e., according to
 $u(r)<a(r)$ or $u(r)>a(r)$. The flow is transonic if at any moment
 it crosses $M=1$. This happens when a subsonic to supersonic or supersonic to
 subsonic transition takes place either continuously or discontinuously.
The point(s) where such crossing
takes place continuously is (are) called sonic point(s),
 and where such crossing takes place discontinuously are called shocks
 or discontinuities.
In order to
satisfy the inner boundary conditions imposed by the
event horizon, accretion onto black holes
exhibit transonic properties in general, which further
indicates that formation of shock waves
are  possible in astrophysical fluid flows onto
galactic and extra-galactic black holes (Das, Pendharkar \& Mitra 2003, and references therein,
Das 2001, and references therein).
The study of dynamical behaviour of such transonic accretion
flow near the BH event horizon is considered to be immensely
important.
Due to the strong curvature 
of space-time close to the black hole, accreting
fluid is expected to show extreme behaviour just before plunging
into the event horizon; this tremendously hot, ultra-fast
flow with it's high-density profile is supposed to provide the
key features of the diagnostic high energy spectra of galactic and extra-galactic
black hole candidates. \\
\noindent
Following the pioneering contributions of Bardeen,
Press \& Teukolsky (1972) and Novikov \& Thorne (1973, NT hereafter),
a number of papers deal with the
general relativistic (GR) BH accretion disc. 
The initial attempt by Fukue (1987 and references therein) and
Chakrabarti (1990, 1996, hereafter C96)
to analytically study the GR transonic BH accretion 
was followed by other independent analytical
works (Kafatos \& Yang 1994,
Yang \& Kafatos 1995, YK hereafter,
Pariev 1996, Peitz \& Appl 1997, Lasota \& Abramowicz 1997, hereafter LA,
Lu, Yu, Yuan \& Young 1997).
%Manmoto 2000, hereafter M00).
Although they made a number of important contributions to the study of
general relativistic {\it advective} black hole accretion discs, none of 
the above mentioned works investigates
the flow
behaviour close to the event horizon  in detail.
Gammie and Popham (Gammie \& Popham 1998,
Popham \& Gammie 1998) performed the only self-consistent analytical work
present in the literature so far which
exploits the 
complete general relativistic formalism to study various flow variables near the event horizon.
Recent interesting work of
Becker \& Le (2003) uses post-Newtonian asymptotic analysis to
study the properties of inner region of
advective accretion flows, and their results are in 
good agreement with general relativistic description. 
However, neither Gammie et al nor Becker et. al. captures the
multi-transonic properties of the flow which
is of considerable astrophysical importance.
Only the multi-transonic flow can produce shock waves
in post-Newtonian 
black hole accretion discs (Das, Pendharkar \& Mitra 2003, and references therein).
Shock waves
in rotating  flows may
provide an efficient mechanism
for conversion of significant amount of the
gravitational energy into
radiation by randomizing the directed infall motion of
the accreting fluid and the hot and dense post-shock flow
is considered to be a powerful tool in
understanding various important astrophysical phenomena like
the spectral properties
of the black hole candidates (Chakrabarti \& Titarchuk 1995, and
references therein),
the formation and dynamics of accretion powered cosmic
jets (Das \& Chakrabarti 1999, and references
therein, Das, Rao \& 
Vadawale 2003), the origin of QPOs in
galactic sources (Das 2003, and references therein).
One thus understands the pressing needs for a 
self-consistent analytical 
model capable of studying the behaviour 
of {\it general relativistic, multi-transonic} accretion flow
{\it sufficiently close} to the event horizon. \\
Motivated by the above mentioned arguments, 
in this paper we concentrate on the stationary, axi-symmetric,
complete general relativistic accretion
solutions  in Schwarzschild metric which contains
multiple critical points, and for such solutions,
we calculate all relevant dynamical
and thermodynamic flow variables {\it extremely close} to the event horizon.
Such a solution scheme for purely spherical general relativistic mono-transonic
BH accretion 
was outlined in Das 2002.
We also exhaustively study the dependence of 
such variables (close to the event horizon) on all important initial
boundary conditions governing the flow.
Our generalized calculation is valid for both
super- as well as for sub-Eddington accretion onto BH
of any mass.
{\it We thus, provide a useful and
self-consistent procedure to study the terminal behaviour
of matter in multi-transonic, general relativistic advective accretion
discs around BHs, which has never been done in any of the
existing works on black hole accretion discs}.
Further calculations, with an ultimate ambition towards
modelling the most general 
viscous, transonic, shocked hydromagnetic accretion in
Kerr-Neumann space time is in progress (T. K. Das, P. J. Wiita \& P. Barai,
in preparation) and would be reported 
elsewhere.\\
\noindent
At this point, we would like to mention that instead of dealing with the so called
standard disc model with significantly high angular momentum (where the presence of
viscous stress allows the infall of accreting material onto the black hole, i.e.,
outward 
viscous transport of angular momentum takes place 
to weaken the centrifugal barrier), we rather
mainly concentrate on accretion flows with relatively low intrinsic angular momentum
(sub-Keplerian angular momentum distribution) where substantially significant
advection velocities may be obtained even for practically inviscid flow, 
although our formalism allows sufficient flexibility to incorporate the viscous 
transonic flows as well, see \S 4.1 for more detail. Such weakly rotating flows
have not been explored much in the literature, although they are well exhibited in
nature for various real physical situations like detached binary systems
fed by accretion from OB stellar winds (Illarionov \&
Sunyaev 1975; Liang \& Nolan 1984), semi-detached low-mass
non-magnetic binaries (Bisikalo et al.\ 1998) and supermassive BHs fed
by accretion from slowly rotating central stellar clusters
(Illarionov 1988; Ho 1999 and references therein). Even for a standerd Keplarian 
accretion disc, turbulence may produce such low angular momentum flow (see,
e.g., Igumenshchev
\& Abramowicz 1999, and references therein).
\section{Formalism}
\subsection{The energy momentum tensor}
\noindent
To provide the most general description of fluid flow in strong
gravity, one needs to solve the equations of motion for the
fluid and the Einstein equations. The problem may be made
simplified by assuming the accretion to be non-self
gravitating so that the fluid dynamics may be dealt in a
metric without back-reactions.
Unless otherwise mentioned for any specific reason(s),
hereafter
we define Schwarzschild radius 
$r_g=\frac{2G{M_{BH}}}{c^2}$, where  $M_{BH}$  is the mass of the black 
hole, $G$ is the Universal gravitational constant and $c$ is the 
velocity of light.
The radial distances and velocities are scaled in units of $r_g$ and $c$
respectively and all other derived quantities are scaled accordingly; 
$G=c=M_{BH}=1$ is used.
We use the Boyer-Lindquist
co-ordinate with signature $-+++$, and an
azimuthally Lorentz boosted orthonormal tetrad basis co-rotating
with the accreting fluid. We define $\lambda$ to be the specific
angular momentum of the flow and neglect any gravo-magneto-viscous 
non-alignment between $\lambda$ and BH spin angular
momentum. \\
\noindent
Let $v_\mu$ be the four velocity of the (perfect) accreting fluid. The energy
momentum tensor $ {\Im}^{{\mu}{\nu}}$ of such a flow could be written as:
$$
{\Im}^{{\mu}{\nu}}=\left(\rho+p\right)v_{\mu}v_{\nu}+pg_{{\mu}{\nu}},
~
{\rm or}~{\bf T}=\left(\rho+p\right){\bf v}{\otimes}{\bf v}+p{\bf g}
\eqno{(1a)}
$$
A complete description of flow behaviour could be obtained
by taking the co-variant derivative of ${\Im}^{{\mu}{\nu}}$
and ${\rho}v^{\mu}$ to obtain the energy momentum
conservation equations and the conservation of baryonic mass.\\
\noindent
However, at this stage, the complete solution remains
analytically untenable unless we are forced 
to adopt a number
of simplified approximations. 
In this paper, as already mentioned in \S 1, 
we would like to study the inviscid
accretion of hydrodynamic fluid in Schwarzchild metric.
Hence, our calculation will be focused on the stationary
axisymmetric solution of the following equations:
$$
{\bf \nabla}^\mu{\Im}^{{\mu}{\nu}}
{\Bigg{\vert}}_{a\rightarrow{0}}=0,~
~
\left({\rho}{v^\mu}\right)_{;\mu}{\Bigg{\vert}}_{a{\rightarrow}0}=0
\eqno{(1b)}
$$
where $a$ is the Kerr parameter,
and the semicolon ($;$) denotes the co-variant derivative.
The radial momentum balance condition 
may be obtained by making the expression
$\left(v_{\mu}v_{\nu}+g_{{\mu}{\nu}}\right){\Im}^{{\mu}{\nu}}_{;{\nu}}$ to be equal to zero.
The `stationarity' condition implies the vanishing of the
temporal derivative of any scalar field in the disc or the
vanishing of the Lie Derivative of any vector or tensor field
along the killing vector $\frac{\partial}{{\partial}t}$, e.g., 
$\frac{{\partial}{\Pi}}{{\partial}r}=0$ if ${\Pi}$ is a scalar
field, and ${\cal L}_{{\partial}/{\partial}t}{\bf {\Pi}}=0$ if
${\bf {\Pi}}$ represents a vector or a tensor field.
The `axisymmetry' property is endowed with the space like
Killing field
$\left(\frac{\partial}{{\partial}\phi}\right)^\mu$.\\
\noindent
Exact Solution of the above mentioned conservation equations
requires the knowledge of accretion
geometry, as well as introduction of a suitable equation of
state. We concentrate on polytropic accretion for which
$p=K{\rho}^{\gamma}$, where $K, \rho$ and $\gamma$
are the monotonic and continuous
function of the specific entropy density, the
adiabatic index and the rest mass density of the flow, respectively.
However, one may note that the polytropic accretion is not the only 
choice to describe the general relativistic transonic black hole accretion and 
equations of state other than the adiabatic one has also been
used by several authors, like YK (isothermal equation)
or Manmoto 2000 (two temperature plasma). There is no reason that more
realistic equation of state can not be employed except the fact that
the analysis would be algebraically more complicated and
hence would be dominated by additional numerical details.
Specific proper flow enthalpy is taken to be 
$h=\left(\gamma-1\right)\left\{{\gamma}-\left(1+a^2\right)\right\}^{-1}$,
where $a$ is the polytropic sound speed defined as (Weinberg 1972, NT, Frank, 
King \& Raine 1992) $a=\left(\partial{p}/\partial{\epsilon}\right)_{\cal S}^{1/2}
={\bf \Psi_1}\left(T(r),\gamma\right)={\bf \Psi_2}\left(p,{\rho},\gamma\right)$,
where $T(r)$ is the local flow temperature,
$\epsilon$ is the mass-energy density,
and $\left\{{\bf \Psi_1},{\bf \Psi_2}\right\}$ are known 
functions. The subscript ${\cal S}$ indicates that the derivative
is taken at constant specific entropy.
\subsection{The disc height}
\noindent
We assume that
the disc has a radius dependent local 
thickness $H(r)$, and it's central plane coincides with
the equatorial plane of the BH.
It is a standard practice 
(Matsumoto et. al. 1984, Paczy'nski 1987, Abramowicz,
Czerny, Lasota \& Szuszkiewicz 1988,
Chen \& Taam 1993,
KY,
Artemova, Bj\"{o}rnsson \& Novikov 1996,
%Narayan, Kato \& Honma 1997,
Wiita 1999,
Hawley \& Krolik 2001, 
Armitage, Reynolds \& Chiang 2001)
to
use the vertically integrated
model in
describing the black hole accretion discs where the equations of motion
apply to the equatorial plane of the BH, assuming the flow to
be in hydrostatic equilibrium in transverse direction.
We follow the same procedure here. Flow
variables, if a vector or tensor, are Lie dragged along
$\left(\frac{\partial}{{\partial}\phi}\right)$
while averaging over all $\phi$, and then are averaged over
a length scale of the order of $H(r)$.
We consider the flow to be
`advective', i.e., to possess considerable radial three-velocity.\\
\noindent
Following the process outlined in LA,
we calculate the disc height.
In
Newtonian framework, the disc height in 
vertical equilibrium is obtained from the $z$ component of the non-relativistic
Euler equation where all the terms involving velocities are neglected and the
higher powers of $\left(\frac{z}{r}\right)$ are ignored in the Taylor's
expansion. For general relativistic disc, the vertical pressure
gradient in the co-moving frame is compensated by the tidal gravitational
field. We 
%drop
%the velocity terms, assume $z<r$ (thin disc approximation),
assign the Lorentz factor $\Gamma$ to be equal to:
$$
\Gamma=\sqrt{\frac{r^3}{r^3-\lambda^2r+\lambda^2}}
\eqno{(2a)}
$$
in the Schwarzschild metric and 
approximate $\frac{\partial{p}}{\partial{z}}\longrightarrow\frac{p}{z}$,
so that the
$z$ component of the 
general relativistic Euler equation provides:
$$
H(r)=2\sqrt{2}\sqrt{\frac{p\left(r^3-\lambda^2r+\lambda^2\right)}{\rho}}
\eqno(2b)
$$
It is trivial to show that for the equation of state and the metric used in this 
paper,
$$
p=\frac{{\rho}a^2\left(\gamma-1\right)}{\gamma^2-\gamma\left(1+a^2\right)}
\eqno(2c)
$$
Using the above equation and the relation (NT, FKR) $a^2(r)=
\frac{\gamma{\kappa}T(r)}{{\mu}m_H}={\Theta}^2T(r)$, where
where $\Theta=\sqrt{\frac{\gamma{\kappa}}{\mu{m_H}}}$,
$\mu$ is the mean molecular weight,
$m_H{\sim}m_p$ is the mass of the hydrogen atom
and $\kappa$ is Boltzmann's constant, we finally obtain the disc height as:
$$
H(r)=5.66{\Theta}T^{\frac{1}{2}}(r)\left[\left(\frac{\gamma-1}{\gamma}\right)
\left\{\frac{r^3-\lambda^2\left(r-1\right)}
{\gamma-\left(1+\Theta^2T(r)\right)}\right\}\right]^{\frac{1}{2}}
\eqno{(2d)}
$$
\subsection{The conserved specific energy of accretion flow}
\noindent
Temporal (zeroth) component of the first part of eq. (1b) leads to
the conservation of specific flow energy ${\cal E}$ 
(relativistic analogue of Bernoulli's constant) along each streamline
as (Anderson 1989) ${\cal E}=hv_t$, where $v_\mu$ is the 
four velocity. 
From the normalization condition of the four velocity ($v_\mu{v^\mu}=-1
$), one writes (NT):
$$
v_t^2=\frac{4r\left(r-1\right)}{\left(1-u^2\right)\left(1-2\Omega{\lambda}\right)
\left(g_{{\phi}\phi}+2\lambda{g_{t{\phi}}}\right)}
\eqno{(3)}
$$
where $\Omega~(=v^{\phi}/v^t)$ is the angular velocity and
$u$ is the radial three velocity in the co-rotating fluid frame.\\
\noindent
One calculates the required metric elements as:
$$
g_{tt}=-\frac{r-1}{r},~g_{{\phi}\phi}=4r^2,~g_{t\phi}=g_{\phi{t}}=0
\eqno{(4a)}
$$
to obtain the expression for the angular velocity as:
$$
\Omega=\left(\frac{r-1}{2r^3}\right)\lambda
\eqno{(4b)}
$$
Hence $v_t$ for a non-spinning (Schwarzschild) black hole may 
be derived as:
$$
v_t{\Bigg{\vert}}_{r_g=\frac{2GM_{BH}}{c^2}}^{\rm Schwarzschild}=
r\sqrt{\frac{r-1}{\left(1-u^2\right)\left(r^3+\lambda^2-\lambda^2r\right)}}
\eqno{(4c)}
$$
Using the value of specific flow enthalpy
$h$ (defined in \S 2.1),
eq. (4c), and $u=aM(r)=\Theta{T^{\frac{1}{2}}(r)}M(r)$, $M(r)$ being the
radial flow Mach number, we obtain the expression for the
conserved specific energy (which includes the rest mass energy)
as:
$$
{\cal E}=\frac{r\left(\gamma-1\right)}
{\left[\gamma-\left\{1+\Theta^2T(r)\right\}\right]}
\sqrt{\frac{r-1}{\left\{1-\Theta^2T(r)M^2(r)\right\}\left(r^3-\lambda^2r
+\lambda^2\right)}}
\eqno{(4d)}
$$
\subsection{The mass and entropy accretion rate}
\noindent
We obtain the mass accretion rate ${\dot M}_{in}$ by integrating the 
second part of the eq. (1b):
$$
{\dot M}_{in}=4{\pi}u(r)
{\sqrt{\frac{r\left(r-1\right)}{1-u^2(r)}}}
{\overline{\Sigma(r)}}_{H(r)}
\eqno{(5a)}
$$
where ${\overline{\Sigma(r)}}_{H(r)}$ is the vertically integrated
surface density of the disc material. Hence one writes:
$$
{\overline{\Sigma(r)}}_{H(r)}={\int_{-{H(r)}/{2}}^{{H(r)}/{2}}}{\rho}(r)dz
\eqno{(5b)}
$$
where ${\rho}(r)$ in is the mean value of radial
density on the equatorial ($z=0$) plane. 
We use the value of $H(r)$ from eq. (2c) to perform the above integration and then
substitute the value of ${\overline{\Sigma(r)}}_{H(r)}$ in
eq. (5a), and use $u(r)=a(r)M(r)=\Theta{T^{\frac{1}{2}}}M(r)$ to 
finally obtain the mass accretion rate as:
$$
{\dot M}_{in}=71.05{\rho}(r){\Theta}T^{\frac{1}{2}}(r)M(r)
\sqrt{\frac{r\left(r-1\right)\left(\gamma-1\right)\left\{r^3-
\lambda^2\left(r-1\right)\right\}}
{\gamma\left\{1-\Theta^2TM^2(r)\right\}\left[\gamma-\left\{1+
\Theta^2T(r)\right\}\right]}}
\eqno{(5c)}
$$
The entropy accretion rate ${\dot {\bf \Xi}}$ can be defined as a quasi-constant 
multiple of the mass accretion rate:
$$
{\dot {\bf \Xi}}
=p^{\frac{1}{\gamma-1}}{\rho}^{\frac{\gamma}{\gamma-1}}{\dot M}_{in}
=K^{\frac{1}{\gamma-1}}{\dot M}_{in}
\eqno{(5d)}
$$
After substituting the value of ${\dot M}_{in}$ from eq. (5c) with the
transformation $T(r)\longrightarrow\frac{a^2(r)}{\Theta^2},~
M(r)=\left\{u(r),a(r)\right\}$, ${\dot {\bf \Xi}}$ reads:
$$
{\dot {\bf \Xi}}=
71.05{\gamma}^{\frac{\gamma+1}{2\left(1-\gamma\right)}}u
\sqrt{
\frac{r\left(r-1\right)\left\{r^3-\lambda^2\left(r-1\right)\right\}}
{1-u^2}}
\left\{\frac{a^2\left(\gamma-1\right)}{\gamma-\left(1+a^2\right)}\right\}
^{{\frac{1}{2}}\left(\frac{\gamma+1}{\gamma-1}\right)}
\eqno{(5e)}
$$
Note that,
in absence of creation and annihilation of matter,
while the mass accretion rate is an absolute constant of motion,
entropy accretion
rate is not. As the expression for ${\dot {\bf \Xi}}$ contains
$K\left(=p{\rho}^\gamma\right)$, which is a measure of the 
entropy density of the flow, ${\dot {\bf \Xi}}$ remains constant 
throughout the flow {\it only if} the local entropy density
of the flow remains unchanged. Thus, ${\dot {\bf \Xi}}$ is a 
constant of motion for shock-free polytropic accretion and 
becomes discontinuous (increases) at the shock location,
if shock forms in the accretion.
Thus for a shock free non-dissipative flow, $\frac{d{\dot {\bf \Xi}}}{dr}=0$
along a streamline. 
\subsection{Dynamical velocity gradient and sonic quantities}
\noindent
We take the logarithmic differentiation (with respect to $r$)
of both the side of eq. (5e) to obtain:
$$
\frac{da(r)}{dr}=
\frac{a}{\gamma+1}\left\{\gamma-\left(1+a^2\right)\right\}
\left[\frac{1}{u\left(u^2-1\right)}\frac{du}{dr}+
{\bf f}\left(r,\lambda\right)\right]
\eqno{(6a)}
$$
where:
$$
{\bf f}\left(r,\lambda\right)=
\frac{1-2r}{2r\left(r-1\right)}+
\frac{\lambda^2-3r^2}{2\left(r^3-\lambda^2r+\lambda^2\right)}
\eqno{(6b)}
$$
We now transform $T(r)\longrightarrow{a(r)},~
M(r)\longrightarrow\left\{u(r),a(r)\right\}$ in the expression
for ${\cal E}$, differentiate both side of eq. (4d) and substitute the 
value of $da(r)/dr$ from eq. (6a) to obtain the dynamical velocity 
gradient as:
$$
\left(\frac{du}{dr}\right)=
\left(
\frac
{
\frac{1}{2r}\left[\frac{2r^3-\lambda^2}{r^3+\lambda^2\left(1-r\right)}
\right]
-\frac{2r-1}{2r\left(r-1\right)}
-\frac{2a^2}{\gamma+1}
\left[\frac{1-2r}{2r\left(r-1\right)}
+\frac{\lambda^2-3r^2}{2\left[r^3+\lambda^2\left(1-r\right)\right]}
\right]
}
{\left[\frac{2a^2}{u\left(u^2-1\right)\left(\gamma+1\right)}
+\frac{u}{1-u^2}\right]}\right)=\frac{{\cal N}}{{\cal D}}
\eqno{(6c)}
$$
Hereafter, we use the notation $\left[{\cal P}_3\right]$ for a set
of values of $\left\{{\cal E},\lambda,\gamma\right\}$, which will be 
our
three parameter initial boundary condition determining the flow behaviour. 
Since the flow is assumed to be smooth everywhere, if
the denominator of eq. (6c)  vanishes at any radial distance
$r$, the numerator must also vanish there to maintain the
continuity of the flow. One therefore arrives at the so
called `sonic point (alternately, the `critical point')
conditions'  by simultaneously making
the numerator and denominator of eq. (6c) equal zero.
The sonic point conditions can be expressed as:
$$
u_s=\sqrt{\frac{2}{\gamma+1}}a_s=
\sqrt{
%\frac{\gamma+1}{2}
\left[
\frac{
\frac{1}{2r}\left[\frac{2r^3-\lambda^2}{r^3+\lambda^2\left(1-\gamma\right)}
\right]
-\frac{2r-1}{2r\left(r-1\right)}
}
{\frac{1-2r}{2r\left(r-1\right)}
+\frac{\lambda^2-3r^2}{2\left[r^3+\lambda^2\left(1-r\right)\right]}
}
\right]_s}
\eqno{(6d)}
$$
where the subscript $s$ indicates that the quantities are to be 
measured
at the sonic point(s). For a fixed $\left[{\cal P}_3\right]$,
we substitute the values of $u_s,a_s$ and $r_s$ in the expression 
of ${\cal E}$ with the substitution $T(r)=a^2(r)/\Theta^2$,
$M(r)=\left\{u(r),a(r)\right\}$, and obtain
the following polynomial, solution of which provides
the sonic point(s) $r_s$:
$$
{\cal  E}^2\left[r_s^3+\lambda^2\left(1-r_s\right)\right]
-\frac{r_s-1}{1-{\bf {{\Psi}\left(r_s,\lambda\right)}}}
\left[\frac{r_s\left(\gamma-1\right)}{\gamma-
{\bf {{\eta}\left(r_s,\lambda\right)}}}\right]^2=0
\eqno{(6e)}
$$
where
$$
{\bf {{\eta}\left(r_s,\lambda\right)}}
=\left[1+\frac{\gamma+1}{2}{\bf 
{{\Psi}\left(r_s,\lambda\right)}}\right]
$$
and
$$
{\bf {{\Psi}\left(r_s,\lambda\right)}}=
\left[
\frac{
\frac{1}{2r_s}\left[\frac{2r_s^3-\lambda^2}{r_s^3+\lambda^2\left(1-r\right)}
\right]
-\frac{2r_s-1}{2r_s\left(r_s-1\right)}
}
{\frac{1-2r_s}{2r\left(r_s-1\right)}
+\frac{\lambda^2-3r_s^2}{2\left[r_s^3+\lambda^2\left(1-r_s\right)\right]}
}
\right]
$$
To
determine the behaviour of the solution near the sonic
point, one needs to evaluate the value of $\left(\frac{du}{dr}\right)$
at that
point (the `critical velocity gradient' $\left(\frac{du}{dr}\right)_s$)
by applying L 'Hospitals' rule to eq. (6c):
$$
\left(\frac{du}{dr}\right)_s={L}t_{r\rightarrow{r_s}}
\frac{\left(\frac{d{\cal N}}{d{r}}\right)}
{\left(\frac{d{\cal D}}{d{r}}\right)}
\eqno{(6f)}
$$
We use the expression of ${\cal N}$ and ${\cal D}$ from eq. (6c),
calculate ${L}t_{r\rightarrow{r_s}}\left(d{\cal N}/dr\right)$
and ${L}t_{r\rightarrow{r_s}}\left(d{\cal D}/dr\right)$.
After a number of complicated algebraic manipulations,
eq. (6f) reduces to the following polynomial which could be solved
to obtain 
the critical velocity gradient:
$$
\frac{2\left(2\gamma-3a_s^2\right)}
{\left(\gamma+1\right)\left(u_s^2-1\right)^2}
\left(\frac{du}{dr}\right)_s^2+
4{\bf {\xi\left(r_s,\lambda\right)}}
\left[\frac{\gamma-\left(1+a_s^2\right)}{u_s^2-1}\right]
\left(\frac{du}{dr}\right)_s
$$
$$
+ \frac{2}{\gamma+1}a_s^2{\bf {\xi\left(r_s,\lambda\right)}}
{\Bigg [}
2{\bf {\xi\left(r_s,\lambda\right)}}\left[
\frac{\gamma-\left(1+a_s^2\right)}{\gamma+1}\right]-
\frac{2r_s-1}{r_s\left(r_s-1\right)}
-\frac{3r_s^2-\lambda^2}{r_s^3+\lambda^2\left(1-r_s\right)}
$$
$$
+\frac{20r_s^3-12r_s^2-2\lambda^2\left(3r_s-2\right)}
{5r_s^4-4r_s^3-\lambda^2\left(3r_s^2-4r_s+1\right)}
{\Bigg  ]}=0
\eqno{(6g)}
$$
where
$$
{\bf {\xi\left(r_s,\lambda\right)}}=
\left[
\frac{1-2r}{2r\left(r-1\right)}
+\frac{\lambda^2-3r^2}{2\left[r^3+\lambda^2\left(1-r\right)\right]}
\right]_{r=r_s}
$$
\section{Results}
\subsection{Parameter space for multi-transonic accretion}
\noindent
For a particular $\left[{\cal P}_3\right]$,
if
${\cal A}\left[{\cal P}_3\right]$ denotes the universal set
representing the {\it entire} parameter space covering all possible
values of $\left[{\cal P}_3\right]$, and if
${\cal B}\left[{\cal P}_3\right]$ represents one particular  subset
of
${\cal A}\left[{\cal P}_3\right]$
which contains  only the
values of $\left[{\cal P}_3\right]$ providing
more than one real roots of eq. (6e),
then ${\cal B}\left[{\cal P}_3\right]$
can further be decomposed into two subsets ${\cal C}\left[{\cal 
P}_3\right]$
and ${\cal D}\left[{\cal P}_3\right]$ such that
$
{\cal C}\left[{\cal P}_3\right]~\subseteq~
{\cal B}\left[{\cal P}_3\right]~
{\rm {for}}~
{\dot {\bf {\Xi}}}\left(r_{in}\right)>
{\dot {\bf {\Xi}}}\left(r_{out}\right),
$
and
$
{\cal D}\left[{\cal P}_3\right]~\subseteq~
{\cal B}\left[{\cal P}_3\right]~
{\rm {for}}~
{\dot {\bf {\Xi}}}\left(r_{in}\right)<
{\dot {\bf {\Xi}}}\left(r_{out}\right),
$
then for $\left[{\cal P}_3\right] \in {\cal C}\left[{\cal P}_3\right]$,
we get {\it multi-transonic accretion} where two real physical inner and outer (with respect to
the BH location) $X$ type sonic points $r_{in}$ and $r_{out}$ encompass
one $O$ type unphysical middle sonic point $r_{mid}$ in between. One can 
obtain multi-transonic {\it winds} for ${\dot {\bf {\Xi}}}\left(r_{in}\right)<
{\dot {\bf {\Xi}}}\left(r_{out}\right)$, e.g., 
$\left[{\cal P}_3\right] \in {\cal D}\left[{\cal P}_3\right]$ provides 
multi-transonic winds.
\\
In Figure 1, the wedge-shaped surfaces represent $
\left({\cal E},\lambda\right) \in \left[{\cal P}_3\right]
\in {\cal C}\left[{\cal P}_3\right]~\subseteq~
{\cal B}\left[{\cal P}_3\right]$
for three different $\gamma=1.3{\dot 3}$ (area bounded by dotted lines), 1.4${\dot 3}$
(area bounded by dashed lines), 1.5${\dot 3}$ (area bounded by solid lines). 
If ${\cal E}_{max}$ be the maximum 
value
of the energy and if $\lambda_{max}$ and $\lambda_{min}$ be the maximum 
and
minimum values of the angular momentum respectively
for ${\cal C}\left[{\cal P}_3\right]$ for a fixed value of 
$\gamma$, 
then
$\left[{\cal E}_{max},\lambda_{max},\lambda_{min}\right]$ non-linearly
anti-correlates with $\gamma$. It is obvious from the 
figure that, as the flow makes a transition
from its ultra-relativistic
\footnote{By the term `ultra-relativistic' and `purely 
non-relativistic'
we mean a flow with
$\gamma=\frac{4}{3}$ and $\gamma=\frac{5}{3}$ respectively,
according to the terminology used in
FKR.} to its purely non-relativistic limit, 
the area representing ${\cal C}\left[{\cal P}_3\right]$
decreases. 
\noindent
\begin{figure}
\vbox{
\vskip -2.5cm
\centerline{
\psfig{file=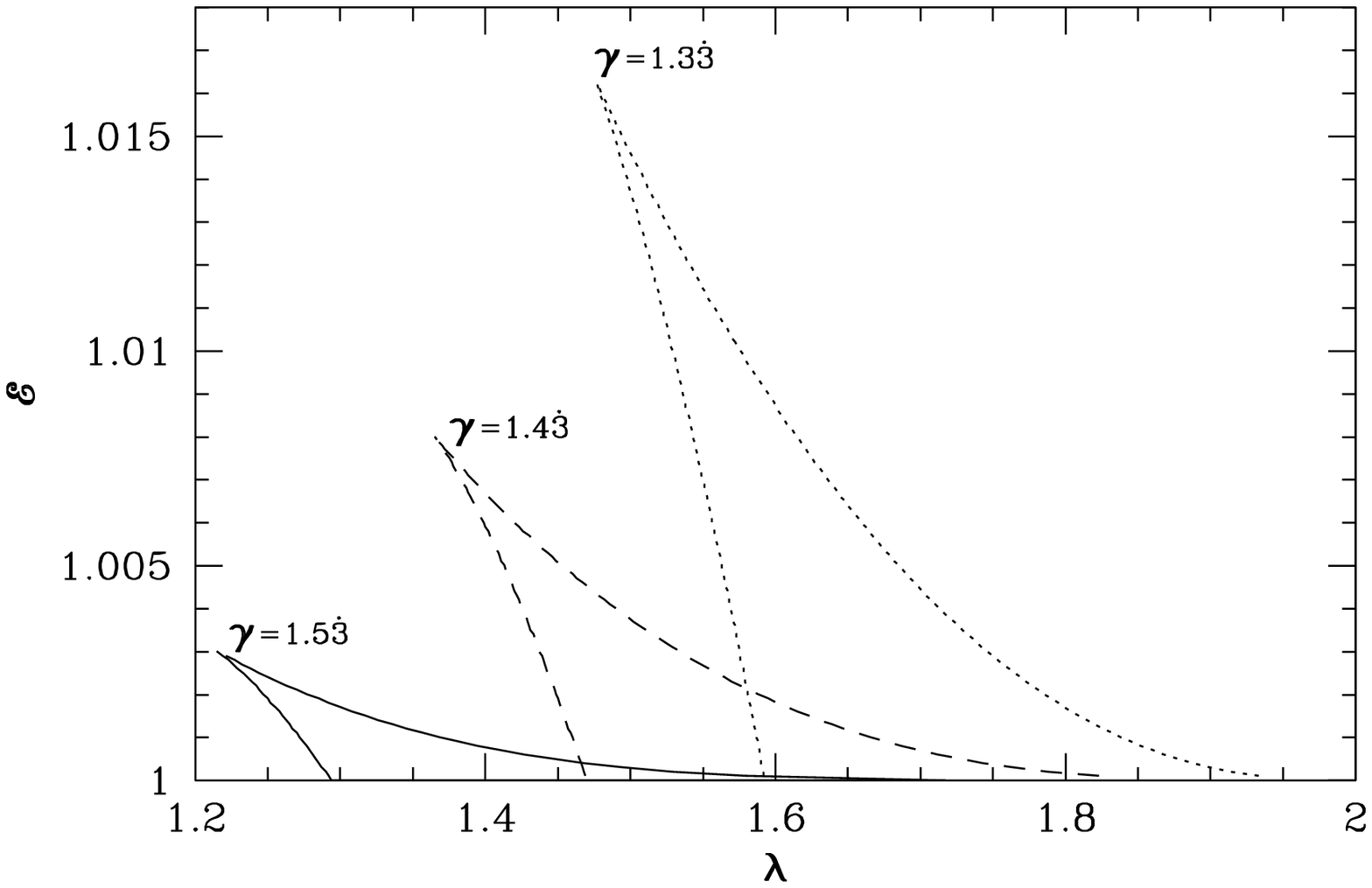,height=20cm,width=17cm,angle=0.0}}}
\noindent {{\bf Fig. 1:}
Parameter space division for multi-transonic black hole accretion.
$\left[{\cal P}_3\right]$ available for such accretion decreases with increase of
$\gamma$, see text for details.}
\end{figure}
Using the formalism developed in \S 2, it is easy to
obtain the similar regions
in $\left[{\cal P}_3\right]$ space representing the multi-transonic
{\it winds} as well. However, as our main interest is to study the 
behaviour of the {\it infalling} matter close to the black hole, we 
do not concentrate on wind solutions in this paper and hence do not 
provide the representative surfaces for $\left[{\cal P}_3\right]{\in}
{\cal D}\left[{\cal P}_3\right]$.
\subsection{Integral curves of motion}
\noindent
In Figure 2, we show the integral curves of motion for multi-transonic
accretion. 
While the distance from the event horizon of the central BH
(scaled in the units of $r_g$ and plotted in logarithmic scale) is
plotted along the X axis, the local  Mach number of the flow is plotted
along the Y axis.
For values of ${\cal C}\left[{\cal P}_3\right]$
shown in the figure, {\bf ABCD}
represents the accretion passing through the outer sonic point {\bf B}, location of which can be found by solving
eq. (6e). {\bf EBI} represents the self-wind. Flow along {\bf GFH} passes
through the inner sonic point {\bf F} and
encompasses a middle sonic point $r_{mid}$ location of which is shown in the figure
using an asterisk. Similar topologies could be obtained for any other
$\left[{\cal P}_3\right]\in{\cal C}\left[{\cal P}_3\right]$. 
The overall scheme for obtaining the above mentioned integral curves is
as follows:\\
\noindent
First we compute $r_{in}$, $r_{mid}$ and $r_{out}$ by solving eq. (6e). Then
we obtain the dynamical velocity gradient of the flow at sonic points
by solving eq. (6g). For a chosen ${\dot M}_{in}$ (scaled in the units of
the Eddington rate ${\dot M}_{Edd}$), we then compute the local dynamical
flow velocity $u(r)$, the local polytropic sound speed $a(r)$, the
local radial Mach number $M(r)$, the local fluid density ${\rho}(r)$
and any other related dynamical or thermodynamic quantities by solving the
eq. (6a-6g) from the outer as well as from the inner sonic point
using fourth order Runge-Kutta method.
Flows passing through
$r_{out}$ may generate ${\Delta}{\dot {\bf \Xi}}={\dot {\bf {\Xi}}}\left(r_{in}\right)
-{\dot {\bf {\Xi}}}\left(r_{out}\right)$ amount of entropy density by 
undergoing
a steady, standing, Rankine-Hugoniot type of shock transition to produce
subsonic, hotter and shock compressed post-shock solutions 
(branch {\bf GFH}) which dives onto the event horizon
supersonically after passing through $r_{in}$. Such shocks are time-like
three-surfaces of first-order discontinuities, and are formed if the following
equation is satisfied:
$$
\Pi{\Sigma}
\left(\frac{T_{-}}{T_{+}}\right)^{\frac{4\gamma}{\gamma-1}}
\left(\frac{1-u_-^2}{1-u_{+}^2}\right)^
{\frac{1}{4}\left(\frac{3-2\gamma}{\gamma-1}\right)}=1
\eqno{(7)}
$$
where $\Sigma(={M_{-}}/{M_{+}})$ and $\Pi(={\dot {\bf \Xi}}_+/{\dot {\bf \Xi}}_-$)
are shock compression and the entropy
enhancement ratio at 
the shock and $T(-/+)$ and $u(-/+)$ are the pre-/post-shock temperature and dynamical
velocities of the flow respectively.
If $\left[{\cal P}_3\right]\in{\cal S}\left[{\cal P}_3\right]$ provides real solutions
of eq. (7), it directly follows that ${\cal S}\left[{\cal P}_3\right]\subseteq
{\cal C}\left[{\cal P}_3\right]$ with
$\left\{\left[{\cal P}_3\right]\in{\cal N}\left[{\cal P}_3\right]
{\big{\vert}}\left[{\cal P}_3\right]\in{\cal C}\left[{\cal P}_3\right], \left[{\cal P}_3\right]\notin{\cal S}\left[{\cal P}_3\right]\right\}$ representing
 the region 
producing {\it non-stationary} shock solutions.
\noindent
\begin{figure}
\vbox{
\vskip -5.5cm
\centerline{
\psfig{file=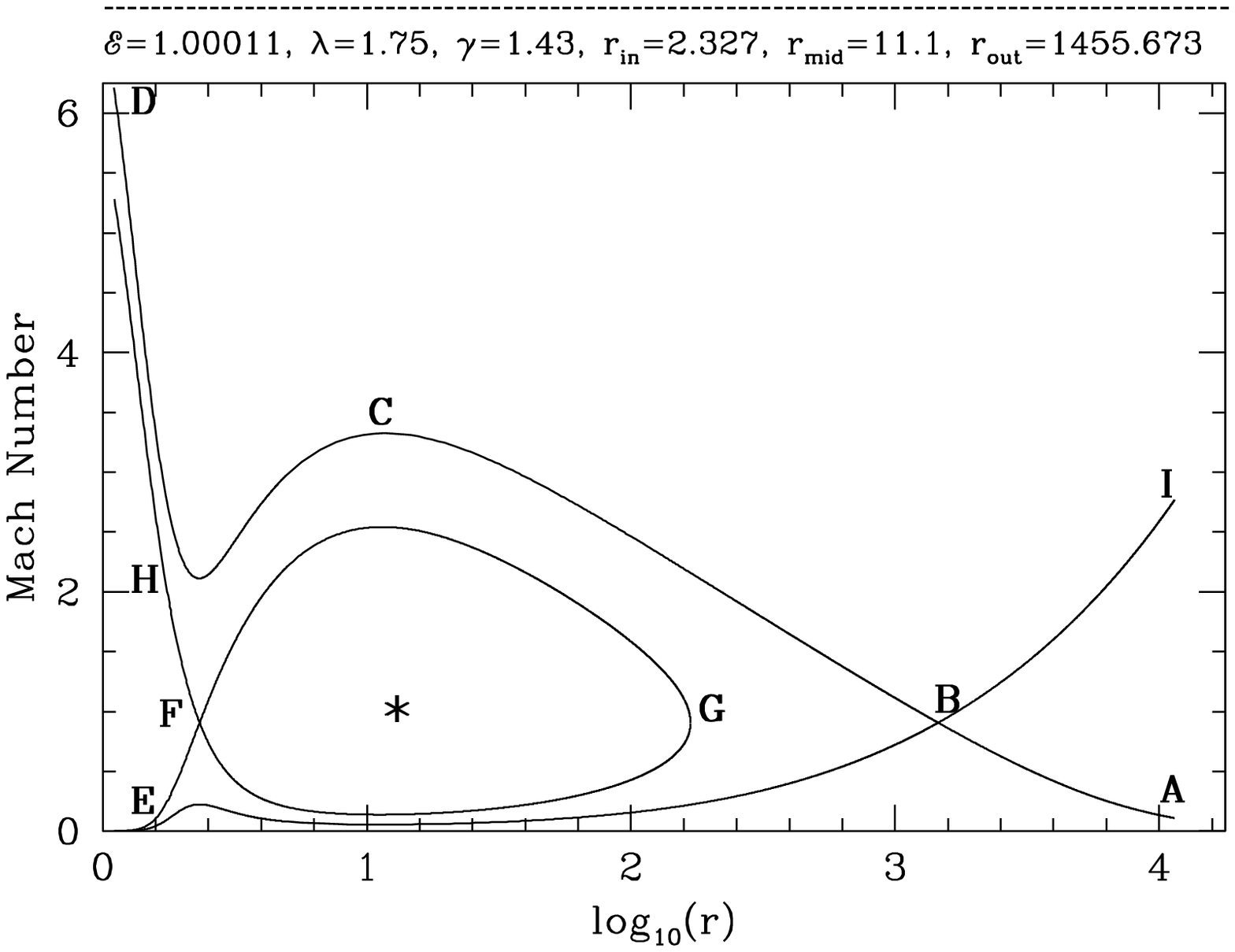,height=19cm,width=17cm,angle=0.0}}}
\noindent {{\bf Fig. 2:}
Solution topology for multi-transonic BH accretion. Sonic parameters (calculated)
and the $\left[{\cal P}_3\right]$ (used) are shown at the top panel of the
figure. See text for detail.
}
\end{figure}
\begin{figure}
\vbox{
\vskip -3.5cm
\centerline{
\psfig{file=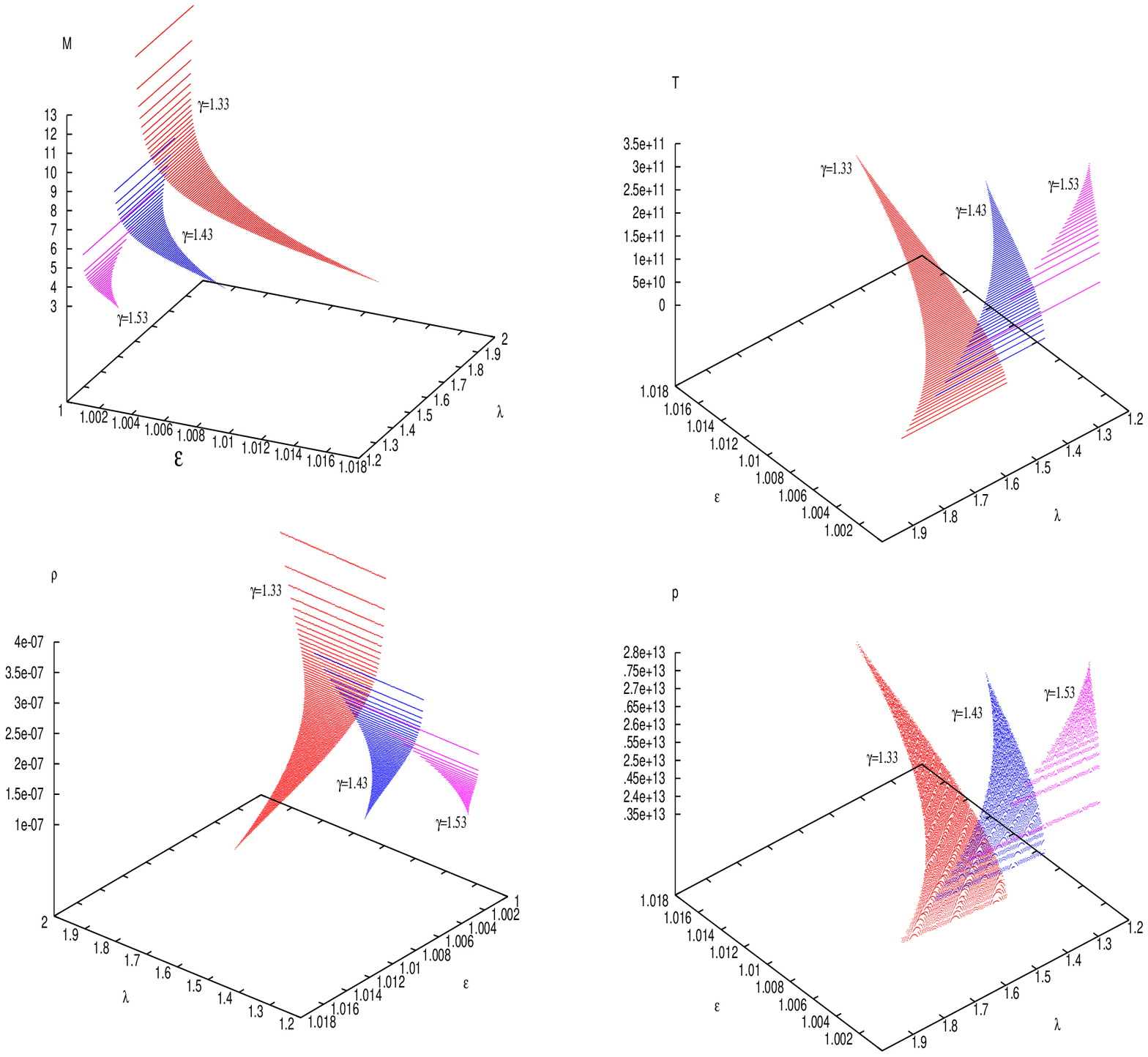,height=26cm,width=23cm,angle=0.0}}}
\noindent {{\bf Fig. 3:}
Quasi-terminal values of Mach number ($M$), temperature ($T$), density ($\rho$) and
pressure ($p$) as a function of $\left\{{\cal E},\lambda\right\}$ for
three different $\gamma$. See text for notes on unit used.
}
\end{figure}
\subsection{Computation of the quasi-terminal values}
\noindent
This section reflects the main findings of our paper.
We define `quasi-terminal values' (QTV) to be the value of any flow variable
$V_f$ as $V^{QTV}_{f}$ at a distance $r_\delta=r_g\left(1+\delta\right)$ with
$0<\delta<<1$. Terminal value ($\delta=0$) of $V_f$ diverges due to singularity at event horizon.
For any $\left[{\cal P}_3\right]\in{\cal C}\left[{\cal P}_3\right]$, we provide 
simultaneous numerical solution of eq. (4d - 6g) to calculate various $V^{QTV}_{f}$.
The basic scheme of such calculation is the following:\\
\noindent
Consider the transonic accretion passing through the outer sonic point $r_{out}$.
We take any $\left[{\cal P}_3\right]\in{\cal C}\left[{\cal P}_3\right]$
and integrate the flow for such $\left[{\cal P}_3\right]$ upto a distance
$\delta$ from the event horizon and calculate the values of various $V_f^{QTV}$,
like $\left\{M(r),T(r),\rho(r),p(r)\right\}^{QTV}$ etc. at $r_\delta$.
We developed an efficient numerical code, which will automatically pick up 
{\it all} $\left[{\cal P}_3\right]\in{\cal C}\left[{\cal P}_3\right]$ 
out of the entire set
$\left[{\cal P}_3\right]\in{\cal A}\left[{\cal P}_3\right]$, and for each
$\left[{\cal P}_3\right]\in{\cal C}\left[{\cal P}_3\right]$, it will
calculate {\it any} $V_f^{QTV}$ at $r_\delta$ for {\it any} small value of 
$\delta$.
For $\delta=0.01$ and $\left[{\cal P}_3\right]_{\gamma=1.3{\dot 3},1.43,1.53}$,
Figure 3 shows the QTV of flow Mach number $M(r)$,
flow temperature $T$ (in degree Kelvin), rest mass flow density $\rho$ (in gm cm$^{-3}$)
and flow pressure $p$ (in dyne cm$^{-2}$) for shock-free accretion passing through
$r_{out}$. $\left\{{\cal E},{\lambda}\right\}$ are plotted
in geometric unit. Any other $V^{QTV}_{f}$ for any $\left[{\cal P}_3\right]$ can be obtained
for $\delta<<0.01$ as well with higher computational cost. Similar calculations can be done for
mono-transonic flows as well. For flows with shocks, general profile of
$V^{QTV}_{f}$ as a function of $\left[{\cal P}_3\right]$ remains fairly 
unaltered with the following difference of the numerical values 
of $V^{QTV}_{f}$ for shocked and shock-free flows:
$$
M^{QTV}_{Shock}<M^{QTV}_{No-Shock},~
\left\{T^{QTV},{\rho}^{QTV},p^{QTV}\right\}_{Shock}>
$$
$$
\left\{T^{QTV},{\rho}^{QTV},p^{QTV}\right\}_{No-Shock}
$$
We observe that the QTV of $u$ at 0.01 $r_g$ may become as much high as 99
percent of the velocity of light, leading to very large values of
flow Mach number near the horizon. The disc thickness in the close proximity of
the event horizon comes out to be 
of the order of few Kilometers. The figure is drawn for  one
Eddington
rate accretion on a 10$M_{\odot}$ BH. Our generalized calculation
allows us to obtain all such $V^{QTV}_f$s for {\it any} accretion rate
onto black holes of {\it any} mass. The following trend is observed for the
variation of $V^{QTV}_f$ with $\left[{\cal P}_3\right]$:
$$
M^{QTV}\propto{\frac{1}{{\cal E}\lambda\gamma}},
u^{QTV}\propto{\frac{\lambda}{{\cal E}\gamma}},
{\rho}^{QTV}\propto{\frac{\lambda}{{\cal E}\gamma}},
T^{QTV}\propto{{\cal E}\lambda\gamma},
$$
$$
p^{QTV}\propto{{\cal E}\lambda\gamma},
H^{QTV}(r)\propto{\frac{\gamma{\cal E}}{\lambda}}
$$
where $\propto$ indicates the proportionality.\\
\noindent
Strongly rotating purely non-relativistic flow with high energy content produces
the maximum disc temperature as well as flow pressure at the event horizon, whereas 
the ultra-relativistic strongly rotating flow with lower energy 
content will produce high density ultra-fast accretion. Disc 
thickness is higher for purely non relativistic weakly rotating 
flows with large value of rest mass energy at infinity, indicating 
the possibility of quasi-spherical inner disc structure
for multiple stellar wind-driven accretion onto black holes. For shocked accretion, the
tendency of forming the quasi-spherical inner disc region is much more
prominent because the entropy generated at shock will puff up the 
post shock disc structure to provide a funnel like surface
which supports the natural
collimation of accretion powered jets (Das \& Chakrabarti 1999). 
\section{Discussions}
\subsection{Predictions for viscous flow}
\noindent
In this work, viscous transport of the angular momentum is not explicitly
taken into account. Even
thirty years after the discovery of standard accretion disc theory
(Shakura \& Sunyaev 1973), exact modeling of viscous multi-transonic
BH accretion, including proper heating and cooling mechanisms is still
quite an arduous task, even for post-Newtonian flow,
let alone for general relativistic accretion. Nevertheless, extremely large radial velocity
close to the BH implies $\tau_{inf}<<\tau_{visc}$ ($\tau_{inf}$ and 
$\tau_{visc}$ are the in fall and the viscous time scales respectively), hence
our assumption of inviscid flow is not at all unjustified, at least upto
a few to tens of $r_g$ or so. Far away from the BH, this 
may not be a very good assumption. However, one understands that
one of the most significant effects of the introduction of
viscosity would be the reduction of radial angular momentum.
We found that the location of the sonic points
anti-correlates with $\lambda$ (weakly rotating flow makes the
dynamical velocity gradient steeper), which indicates that for 
viscous flow the sonic points will be pushed further out and the flow would
become supersonic at a larger distance for the same set of other initial
boundary conditions. The terminal values of the Mach number and disc height
anti-correlate, while the density and pressure
correlates with $\lambda$. Our calculations, if applied for viscous flow,
will thus produce a more supersonic and quasi-spherical structure near the event horizon,
while the terminal flow pressure and density would be reduced.
\subsection{Spectral Signature of Black Hole Spin}
\noindent
One of the most tantalizing issues in black hole 
astrophysics is to study whether
the spin of the black hole 
(the Kerr parameter $a$) could be determined using any 
observational means, the spectral signature of the black hole, for example.
Our investigation of general relativistic
accretion flows in Kerr metric may provide an important 
step towards a better understanding of this problem. For accretion onto
the Schwarzschild black holes, we have been able to track the infalling
matter upto the point extremely close to the event horizon. We can
successfully perform the same procedure (with increasing degree of
mathematical complexity) for accretion onto Kerr black hole
as well and 
can calculate various dynamical and thermodynamic flow variables
upto the point extremely close to the event horizon (as close as it could be)
as a function of Kerr parameter $a$ (T. K. Das,
Paul J. Wiita \& Paramita Barai, in preparation). 
Hence the terminal behaviour of
infalling matter (e.g., flow temperature, pressure, density
etc. which are responsible to characterize the observed spectra) 
could be studied as a function of the black hole
spin and the spectral signature 
of the black hole
rotation may, at least qualitatively, be determined. This,
will, as we believe, be an important step forward towards
a better understanding of black hole astrophysics. 
However, we would like to mention here that such spectral
signatures would be very difficult to detect observationally. Most of the
radiations produced such a close vicinity of the event horizon will be
directly swallowed by the black hole itself, rest of the radiative
signature will eventually suffer an enormous amount of gravitational 
redshift, and hence will remain almost inaccessible for analysis using
any present day observational techniques. Also, the temperature profile obtained from
our work will mainly correspond to the extreme ultra-violet or X-rays, while most of the
{\it observed} flux for black hole candidates come from flares and jets instead of the
inner region of the disc, and hence the theoretically calculated
strong gravity effects would be diluted.
\subsection{Hawking radiation from acoustic black holes}
\noindent
The Pioneering attempt made by Unruh (1981) on mapping certain aspects 
of BH physics onto the propagation of sound wave in supersonic 
convergent flows leads to a number of important work (see Visser 1998, V98 hereafter, and
references
therein)
to identify the propagation of acoustic disturbance in a 
polytropic inviscid fluid flow with the d'Alembertian equation of
motion of a minimally coupled massless scalar field propagating in a
(3+1) Lorentzian manifold. Sound propagation is described by an
acoustic metric algebraically dependent on the flow density and Mach
number; which is conformaly related to the 
Painleve$^{'}$-Gullstrand-Lemaitre
representation of Schwarzschild geometry, with a constant conformal
factor close to the black hole event horizon. Such an equivalence 
is potentially useful to tackle the tantalizing issue of
Hawking radiation in terms of physical quantities governing the 
transonic fluid flow. 
In this paper we construct
multiple sonic surfaces
(a collection of $r_{in}$ and $r_{out}$ forming a
3D hypersurface on
$\left[{\cal P}_3\right]$ space) which is
essential for a complete investigation of the formation of
outer-trapped surfaces in acoustic ergo-region and of acoustic
horizons.
We investigate the multi-transonic,
barotropic inviscid fluid flow (accretion) using a {\it complete}
general relativistic framework and {\it not} using any Newtonian assumption.
Hence, we
believe that our work presented in this letter, may have a
two fold importance. Firstly, a general relativistic calculation of the values of the
fundamental dynamical and thermodynamic flow variables close to the
black hole event horizon will help in constructing a diagnostic high energy spectra,
useful for the identification of galactic and extragalactic black hole candidates.
Secondly, a full general relativistic treatment of representative transonic fluid flow
will take an welcome step-forward towards a concrete formulation of the
`Acoustic General Relativity', as the term used by V98.
We have also established a very important fact that general relativistic spherical 
accretion onto astrophysical black holes may generate acoustic {\it white holes}, 
and the ratio of the Hawking temperature to the
analogous sonic black hole temperature is {\it independent} of the black hole mass,
which is a nice way out from  the great debate whether astrophysical/ primordial
black holes can really have such a lower mass limit so that the black hole
micro-physics can really be dealt with. Details of such calculations are beyond the scope of
this paper and have been presented elsewhere.
\section{Acknowledgments}
\noindent
This work
is supported by Grant No. NSF AST-0098670.
It is my pleasure to acknowledge stimulating discussions with Paul J. Wiita,
Robert V. Wagoner, William Unruh and Paramita Barai. I would also like to thank the
anonymous referee for useful comments.
{}
\end{document}